\documentclass[a4paper,twoside,10pt]{letter}
\usepackage{graphicx,saj,multicol,subeqnarray}
\usepackage{url}

\def\papertype{Original Scientific Paper}

\def\udc{...}
\begin{document}
\baselineskip=3.1truemm
\columnsep=.5truecm
\newenvironment{lefteqnarray}{\arraycolsep=0pt\begin{eqnarray}}
{\end{eqnarray}\protect\aftergroup\ignorespaces}
\newenvironment{lefteqnarray*}{\arraycolsep=0pt\begin{eqnarray*}}
{\end{eqnarray*}\protect\aftergroup\ignorespaces}
\newenvironment{leftsubeqnarray}{\arraycolsep=0pt\begin{subeqnarray}}
{\end{subeqnarray}\protect\aftergroup\ignorespaces}
%


\markboth{\eightrm CONTRIBUTION OF L\lowercase{Y}$\alpha$ LINE IN IONIZATION RATE IN D-REGION} {\eightrm A.
NINA, V. M. {\v C}ADE{\v Z} and J. BAJ{\v C}ETI{\' C}}

{\ }

\publ

\type

{\ }


\title{Contribution of Solar Hydrogen L\lowercase{Y}$\alpha$ line emission in total ionization rate in ionospheric D-region during the maximum of solar X-flare }


\authors{
        A. Nina$^{1}$,
        V. M. \v{C}ade\v{z}$^{2}$
        and J. Baj\v{c}eti\'{c}$^{3}$}

\vskip3mm


\address{$^1$Institute of Physics,
University of Belgrade\break Pregrevica 118, 11080 Belgrade,
Serbia}

\Email{sandrast}{ipb.ac.rs}

\address{$^2$Astronomical Observatory, Volgina 7, 11060 Belgrade, Serbia}

\Email{vcadez}{aob.rs}

\address{$^3$Department of Telecommunications and Information Science, University of Defence, Military Academy, Generala Pavla Juri\v{s}i\'{c}a \v{S}turma 33, 11000 Belgrade, Serbia}

\Email{bajce05}{gmail.com}


\dates{--}{--}


\summary{The solar Ly$\alpha$ line emission can be considered as the dominant source of ionization processes in the ionospheric D-region at altitudes above 70 km during unperturbed conditions. However, large sudden impacts of radiation in some other energy domains can also significantly influence the ionization rate and, in this paper, we present a study on the contribution of Ly$\alpha$ radiation to the ionization rate when the ionosphere is disturbed by solar X-flares. We give relevant analytical expressions and make calculations and numerical simulations for the low ionosphere using data collected by the VLF receiver located in Serbia for the VLF radio signal emitted by the DHO transmitter in Germany. }


\keywords{solar-terrestrial relations -- Sun: activity -- Sun: flares -- Sun: X-rays,
gamma rays.}

\begin{multicols}{2}
{


\section{1. INTRODUCTION}

A specific property of the ionospheric medium located in the terrestrial atmosphere at altitudes between around 50 km and 1000 km is presence of charged particles which play a significant role in numerous physical and chemical processes and have influence on various natural and man induced features occurring in this area. Consequently, the investigations of particle densities in the ionosphere and particularly in its lowest part which is in focus of our investigations, are very important for scientific studies such as analyses of plasma parameters changes induced by solar X-flares (Nina et al. 2011, Schmitter 2013), solar eclipses (Sing et al. 2011), lightnings (Inan et al. 1988), or induced harmonic and quasi-harmonic hydrodynamic motions (including soliton formation and vortices)  (Jilani et al. 2013, Nina and \v{C}ade\v{z} 2013, Maurya et~al. 2014, Zhang and Tang 2015). There is also a significance of  practical applications in telecommunications as the low ionosphere electron density variations have the greatest effect on radio wave propagation, primarily within the low (30kHz-300kHz),very low (3kHz-30kHz) and ultra-low (0.3kHz-3kHz) frequency bands. Prediction and forecasts of such events is important for variety of reasons including the remote sensing detection of narrow bipolar events in clouds (Ushio et al. 2014), monitoring acoustic and gravity waves in the atmosphere (Nina and \v{C}ade\v{z} 2013), monitoring specific perturbations in the ionosphere that might precede seismic activity (Hayakawa et al. 2010) and tropical cyclones (Price et al. 2007), and analysis of radio signal propagation characteristics (Baj\v{c}eti\'{c}} et al. 2015).

Although the low ionosphere is permanently under different ionizing radiation influences, some of them dominate under the considered conditions. Thus, the ionization of its lowest daytime region, the D-region (60 km - 90 km altitude), is primary caused by the Ly$\alpha$ radiation from the solar hydrogen. However, intensive events can significantly perturb this part of the atmosphere. For example, solar X-flares can induced large electron density increase in daytime ionosphere (\v{S}uli\'{c} and Sre\'{c}kovi\'{c} 2014). This event make the contributions of relevant parts of electromagnetic spectrum in ionization processes time dependent which may change the dominant source of electron productions. Also, contrary to the most important influence of electromagnetic waves in electron production during daytime conditions, the ionization of nighttime low ionosphere is controlled by the eventual high energy corpuscular particles (\v{Z}igman et al. 2014).

One of the most important sudden perturbation source is the increase of the X radiation from the Sun during solar X-flares. In some of these cases when the flux of photons in the X domain is sufficiently large, this radiation dominates in ionization processes of the D-region. In literature, this is usually considered as approximately the total ionization during the time period around the maximum X-radiation intensity (Ratcliffe 1972, Budden 1988, {\v Z}igman et al. 2007). However, there are no quantitative analyses of contribution of Ly$\alpha$ photons in electron production based on data obtained by monitoring a particular part of the D-region during considered time period. Consequently, there are no relevant procedures found in the literature that would justify the exclusion of Ly$\alpha$ photons from calculations in such cases.

In this paper our attention is focused on contribution of the Ly$\alpha$ emission in ionization of the ionospheric D-region at the time around the maximum of X radiation during a solar X-flare, and on estimates of its relative significance depending on altitude.
The derived theoretical equations are applied to a particular case of the D-region perturbation induced by a solar X-flare occurred on May 5, 2010.
For the low ionosphere observation we used the method based on very low frequency (VLF) radio signals as also done in numerous published studies (Clilverd et al. 1999, Inan et al. 2010, Kolarski et al. 2011, Nina et al. 2012a,b, Singh et~al. 2014).

\section{2. EXPERIMENTAL SETUP AND OBSERVED DATA}
\label{exp}

The low ionosphere is monitored by three suitable techniques based on very low frequency (VLF) radio waves, rocket, and radar measurements (see for example Grubor et al. 2005, Strelnikova and Rapp 2010 and Chau et al. 2014).
The first of them, which is used in this work, is based on continuously emitted and recorded radio signals by numerous, worldwide distributed, transmitters and receivers.

\vskip1.7cm

\centerline{\includegraphics[bb=0 0 629
815,width=\columnwidth, keepaspectratio]{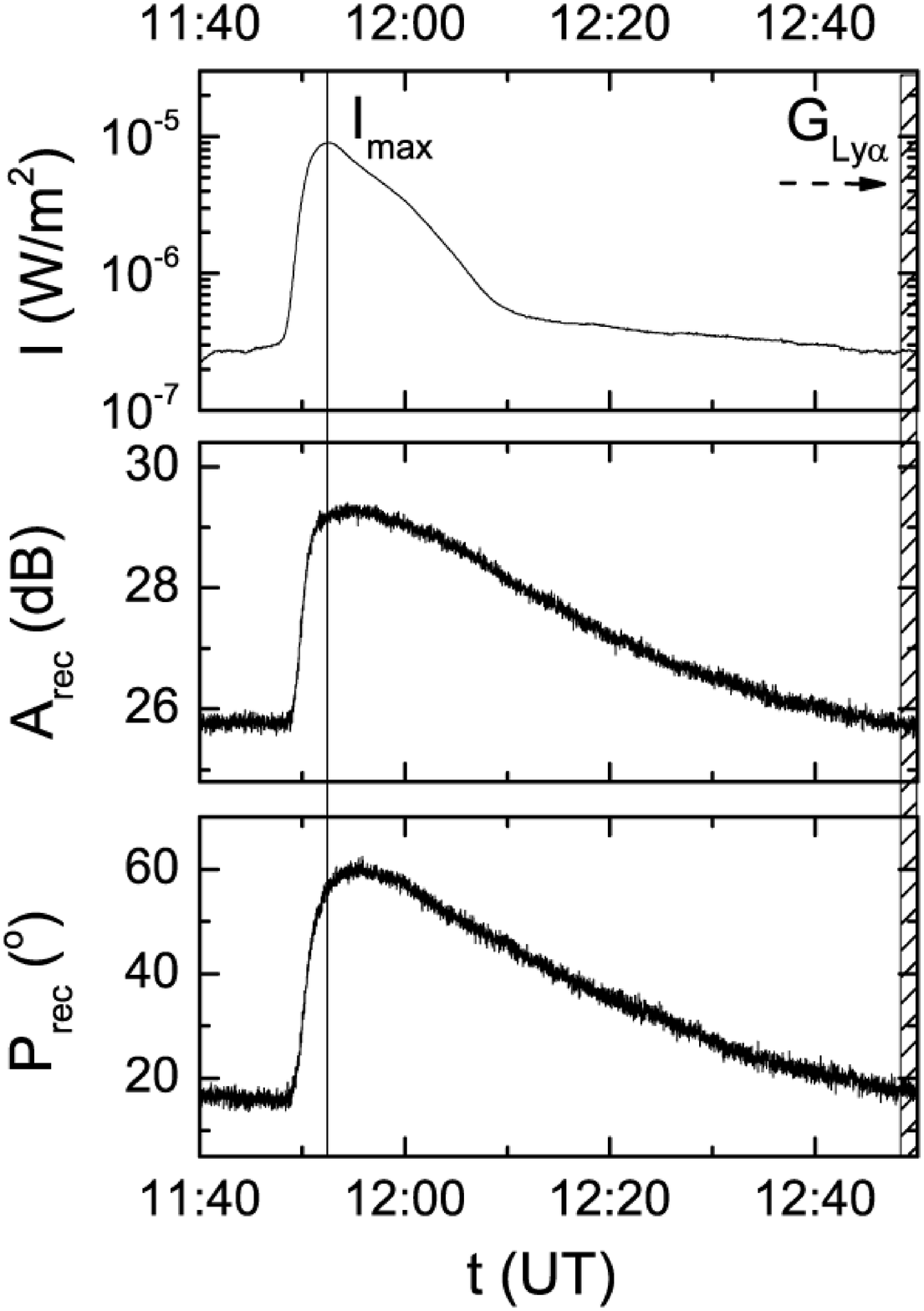}}

\figurecaption{1.}{Increase of X-radiation registered by the GOES-14 satellite on May 5, 2010 in wavelengths domain 0.1 nm - 0.8 nm (upper panel), and reaction of amplitude (middle panel) and phase (bottom panel) of the VLF signal emitted by the DHO transmitter located in Germany and received by the AWESOME receiver in Serbia. The vertical line indicates the time of X radiation maximum and the shaded domain represents the relaxation period analyzed in calculation of ${\cal G}_{Ly\alpha}$ in Nina and {\v C}ade{\v z} 2014.}

Our analysis is based on data obtained from the low ionosphere monitoring using the 23.4 kHz VLF signal emitted by the DHO transmitter in Rhauderfehn (Germany) and received in Belgrade (Serbia). This transmitter was chosen because it provides the best quality
of the recorded signal due to its high emission power of 800 kW and
suitable signal frequency for the location of the receiver, and a relatively short signal propagation path.
The latter property is important as it excludes significant variations in vertical stratification of parameters
in the ambient ionospheric plasma.

The final theoretical results of the study and numerical procedure for modelling the D-region plasma are applied to a case of perturbation induced by the solar X-flare occurred on May 5, 2010 with the X radiation intensity $I$ registered by the National Oceanic and Atmospheric Administration (NOAA) satellite GOES-14, (Fig. 1,
 the upper panel) at wavelengths range 0.1 nm - 0.8 nm. The ionosphere perturbations were detected as amplitude $\Delta A _{\rm {rec}}$ and phase $\Delta P _{\rm {rec}}$ variations of the considered VLF signal recorded by the AWESOME (Atmospheric Weather Electromagnetic System for Observation Modelling and Education) VLF receiver (Cohen et~al. 2010) (bottom and middle panels, respectively). In Fig. 1, the time evolutions of recorded data are shown for the whole perturbation period. However, the presented study is related to the time of the X-radiation intensity maximum $t_{\rm {I_{max}}} = 11:52:40$ UT indicated by the vertical line while the shaded domain designates the period considered in calculation of the Ly$\alpha$ photoionization rate ${\cal G}_{{\rm Ly\alpha}}$ as presented in Nina and {\v C}ade{\v z} 2014.

\section{3. D-REGION MODELLING}
\label{modelling}

The electron gain rates induced by the Ly$\alpha$ line and X radiation, ${\cal G}_{\rm {_{\rm {Ly\alpha}}}}$ and ${\cal G}_{\rm X}$, respectively, during the time of the X radiation maximum can be calculated from the equation for the D-region electron density dynamic (McEwan and Phillips 1978):

\begin{equation}
\label{eq:dne}
    {
    \frac{dN_{\rm e}(\vec{r},t)}{dt}= {\cal G}(\vec{r},t)-\alpha_{{\rm eff}}(\vec{r},t)N_{\rm e}^{2}(\vec{r},t).
    }
\end{equation}
where the influence of transport processes is neglected as they become important only
  at altitudes above 120-150 km (Blaunstein and Christodoulou 2006). 
The total electron gain rate ${\cal G}$ can be written as:

\begin{equation}
\label{eq:g}
    {
    {\cal G}(\vec{r},t)={\cal G}_{_{\rm {Ly\alpha}}}(\vec{r},t)+{\cal G}_{\rm {X}}(\vec{r},t).
    }
\end{equation}

The satellite observations show that increasing of the $Ly\alpha$ radiation is not always recorded during periods of solar X-flares (Raulin et al. 2013). In the cases without important changes in the Ly$\alpha$ line intensity it can be taken that ${\cal G}_{{\rm Ly\alpha}}$ does not vary significantly in time if the considered time period is sufficiently short. Thus, the total electron gain rate can be taken constant in time around the X radiation intensity maximum ($d{\cal G}_{\rm X}(\vec{r},t)/dt=0$). At the same conditions, also the coefficient $\alpha_{{\rm eff}}(\vec{r},t)$ can be considered practically stationary ({\v Z}igman et~al. 2007).
Using these approximations, the time derivative of Eq. (\ref{eq:dne}) gives:

\begin{equation}
\label{eq:d2ne}
    {
    \frac{d^2 N_{\rm e}(\vec{r},t)}{dt^2}=-\alpha_{{\rm eff}}(\vec{r},t_{_{\rm {I_{max}}}})\frac{dN_{\rm e}^{2}(\vec{r},t)}{dt}.
    }
\end{equation}
Finally, Eqs. \ref{eq:dne} and \ref{eq:d2ne} yield the following expression for ${\cal G}(\vec{r},t_{_{\rm {I_{max}}}})$:

\begin{equation}
\label{eq:gImax}
    {
    \begin{array}{ll}
    {\cal G}(\vec{r},t_{_{\rm {I_{max}}}})= \left.\frac{dN_{\rm e}(\vec{r},t)}{dt}\right\vert_{t=t_{I_{max}}} \\[.5cm]- \left.\frac{d^2 N_{\rm e}(\vec{r},t)}{dt^2}\left[\frac{dN_{\rm e}^{2}(\vec{r},t)}{dt}\right]^{-1}N_{\rm e}^{2}(\vec{r},t)\right\vert_{t=t_{{\rm I_{max}}}}.
    \end{array}
    }
\end{equation}

The procedure to calculate ${\cal G}_{{\rm Ly\alpha}}(\vec{r},t)$ is presented in Nina and {\v C}ade{\v z} 2014,
 while the electron density time evolution and, consequently, its time derivatives can be obtained from comparisons of the recorded amplitude $\Delta A _{\rm {rec}}$ and phase $\Delta P _{\rm{rec}}$ changes with the corresponding values resulting from the LWPC (Long-Wave Propagation Capability) numerical modeling the VLF signal propagation (Ferguson 1998) as explained in Grubor et al. 2008. The procedure is based on finding the combination of input parameters, the signal reflection height $H'$ (in km) and sharpness $\beta$ (in km$^{-1}$), that gives the best matching of the recorded and modeled signal characteristics. The electron density $N_{\rm e}(h,t)$ (in m$^{-3}$) at fixed altitude $h$ (in km) is calculated from these parameters by applying Wait's model of the ionosphere considering the vertical electron density gradient (Wait and Spies 1964):

\begin{equation}
\label{eq::Ne}
N_{\rm e}(h,t) = 1.43\cdot10^{13}e^{-\beta(t)H'(t)}e^{(\beta(t)-0.15)h}.
\end{equation}

Finally, Eq. \ref{eq:gImax} and the obtained values for $N_{\rm e}$ from Eq. \ref{eq::Ne} yield the percentage contribution $r_{_{\rm{Ly\alpha}}}(h)=100{\cal G}_{_{\rm{Ly\alpha}}}(h)/{\cal G}(h)$
 of the Ly$\alpha$ line in the ionization rate at time of the maximum X radiation intensity as follows:

\begin{equation}
    \label{eq:13a}
    {
\begin{array}{ll}
 r_{_{\rm{Ly\alpha}}}(h)=100{\cal G}_{_{\rm{Ly\alpha}}}(h) \\[.5cm] \left.\left [\frac{dN_{\rm e}(h,t)}{dt}
-\frac{d^2 N_{\rm e}(h,t)}{dt^2}\left[\frac{dN_{\rm e}^{2}(h,t)}{dt}\right]^{-1}N_{\rm e}^{2}(h,t) \right ]^{-1}\right\vert_{t=t_{\rm {I_{max}}}}.
\end{array}
    }
  \end{equation}

\section{4. RESULTS AND DISCUSSION}

As already said in Section 2, the obtained equation for the fractional contribution of the Ly$\alpha$ line in ionization rate at the maximum X radiation intensity and the needed characteristics of electron density are applied to the time period of the ionospheric perturbation induced by the solar X-flare from May 5, 2010. To examine the validity of approximation of stationary Ly$\alpha$ radiation photoionization rate
as assumed in the procedure given in Section 3, we compare rises of the X and Ly$\alpha$ radiation intensity using the model that determines radiation intensity spectra  $I_{\rm {\lambda}}$
 during solar flares given in Chamberlin et~al. 2008. Fractional increases  of the X and Ly$\alpha$ radiation intensity are presented by the parameter  $b_{\rm {\lambda}}$
\begin{equation}
\label{eq::b}
b_{\rm {\lambda}}=100\frac{(I_{\rm {\lambda}}(t_{ \rm {Imax}})-I_{\rm {\lambda}}(t_{ \rm {up}}))}{I_{\rm {\lambda}}(t_{ \rm {up}})},
\end{equation}
where $t_{ \rm {Imax}}$ and $t_{ \rm {up}}$ designate times of the flare maximum and unperturbed time period before X-flare, respectively.

\vskip1cm

\centerline{\includegraphics[bb=0 0 629
815,width=\columnwidth, keepaspectratio]{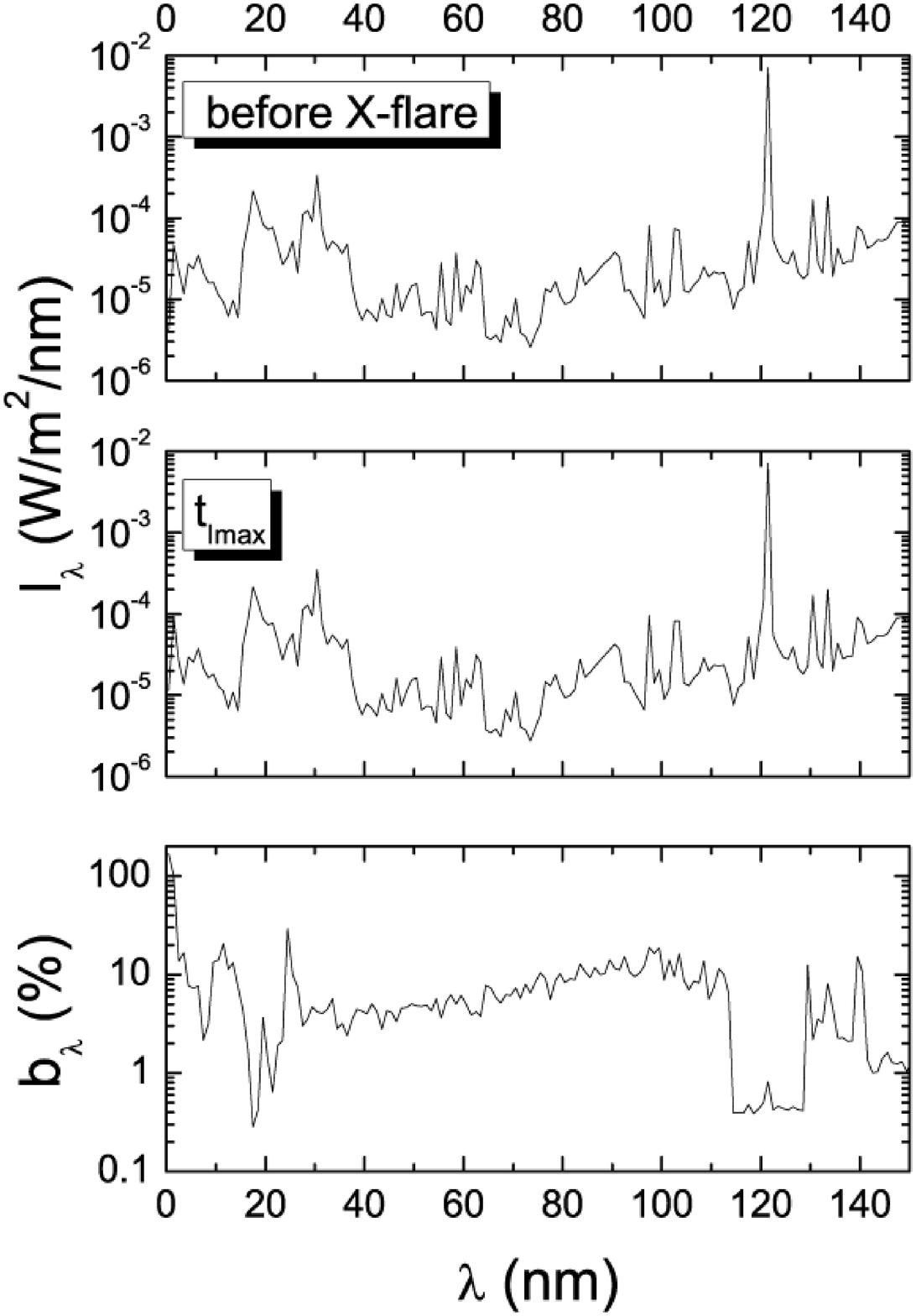}}

\figurecaption{2.}{Radiation spectra obtained by the model given in Chamberlin et~al. 2008 at time periods before the X-radiation flare on May 5, 2010 (upper panel) and at the  maximum X-radiation intensity (middle panel). The increasing $b_{\rm {\lambda}}$ of relevant values are given in the bottom panel.}

The considered spectra shown in Fig. 2 (upper and middle panels) and the obtained coefficient $b_{\rm {\lambda}}$ plotted in the bottom panel of this figure show that variations of Ly$\alpha$ radiation are negligible. The obtained values for the X radiation within domains of wavelength between 0 nm and 1 nm (domain of X radiation observed by GOES satellite (Fig. 1) fall within this domain), and between 120 nm and 121 nm (Ly$\alpha$ photons fall within this domain) are around 170\% and below 1\% which shows that variations in Ly$\alpha$ radiation are negligible in comparison with the X radiation. Here we want to point out that the Ly$\alpha$ photons can ionize only NO molecules in the D-region. During solar X-flares the X radiation decreases the density of this molecule which, consequently, additionally decreases the ${\cal G}_{_{\rm{Ly\alpha}}}$. So, the real contribution of this line in total ionization is lower then what we obtained. The study of this difference will be subject of our future investigations.

Fig. 3, shows the calculated electron density evolutions $N_{\rm e}$ for indicated time periods at altitudes 70 km, 72 km, 74 km, 76 km, 78 km and 80 km. The obtained values are in good agreements with those shown in {\v Z}igman et~al. 2007, Grubor et al. 2008 and Kolarski and Grubor 2014. Because of numerous simultaneously active influences, the calculated quantities cannot be smooth functions of time $t$. To extract the dominant influence of solar flare and to find a smooth time derivative of electron density we fitted the given discrete points with second order polynomials as shown in Fig. 3.

 \vskip-4.7cm

 \centerline{\includegraphics[bb=0 0 629
815,width=\columnwidth, keepaspectratio]{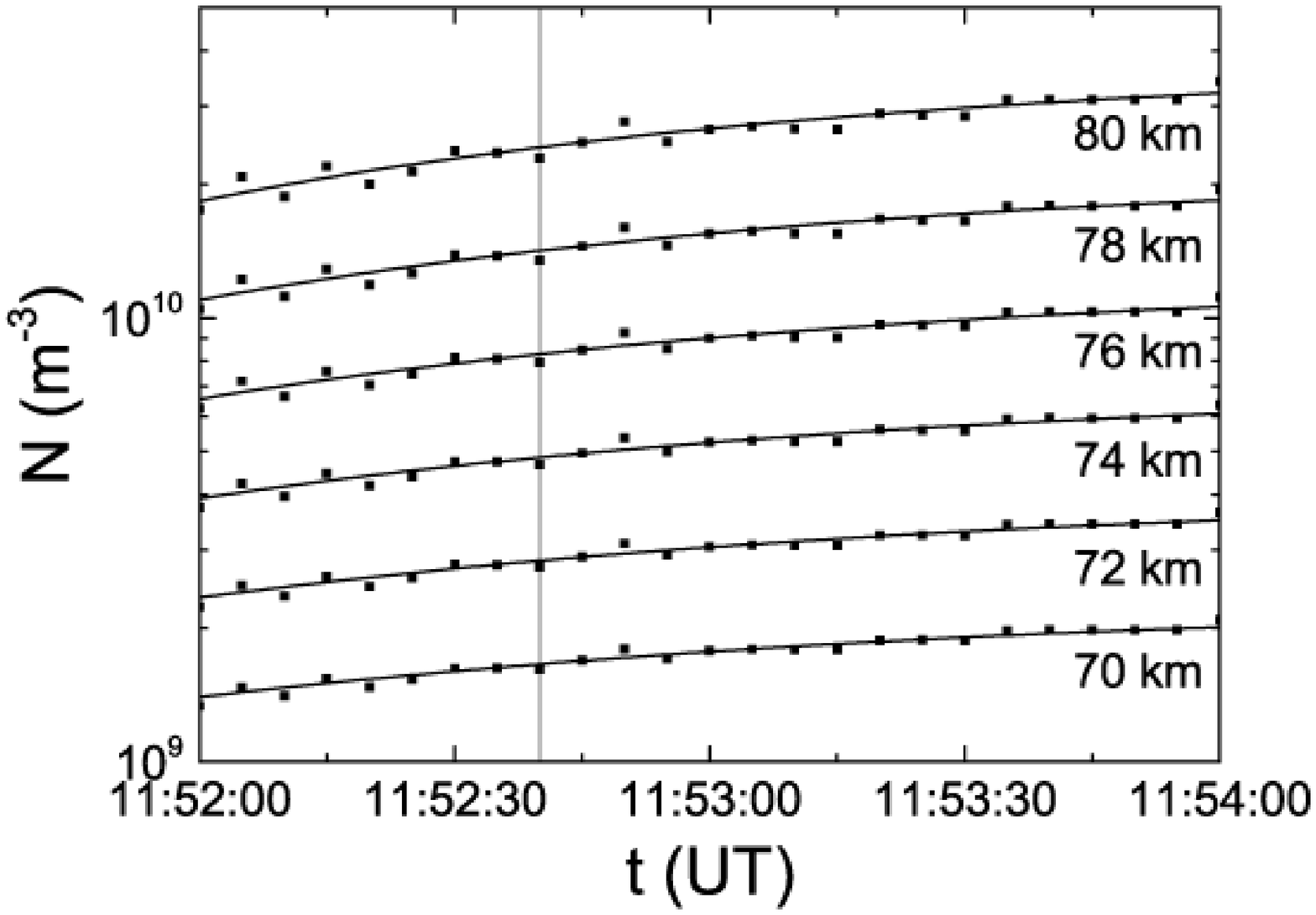}}

\figurecaption{3.}{Electron density time evolutions around the X radiation maximum ($t_{\rm {I_{max}}}$) for altitudes 70 km, 72 km, 74 km, 76 km, 78 km and 80 km.}

The first and second time derivative of electron density are shown in the upper panel in Fig. 4. It can be seen that the first derivative is positive while the second derivative takes negative values. For both of them, their absolute values increase with altitude indicating a larger influence of solar flares at higher altitudes.

Fig. 5 shows the resulting altitude dependencies of the relative contribution $r_{\rm {Ly\alpha}}$ of the Ly$\alpha$ line in the ionization rate at the time of maximum X radiation intensity calculated from Eq. \ref{eq:13a}. We can see that this coefficient has values below 0.5\% in the considered case. This justifies the assumption that the X radiation (including spectral lines and the continuum) dominate in ionization processes within the D-region during the period around the X radiation intensity maximum, as taken in literature (see for example {\v Z}igman et~al. 2007). It can also be seen that the significance of the Ly$\alpha$ radiation in ionization processes decreases with altitude under the considered conditions. In the case of the analyzed flare, our procedure gives values between around 0.1\% and 0.4\%.

 \vskip1.7cm

\centerline{\includegraphics[bb=0 0 629
815,width=\columnwidth, keepaspectratio]{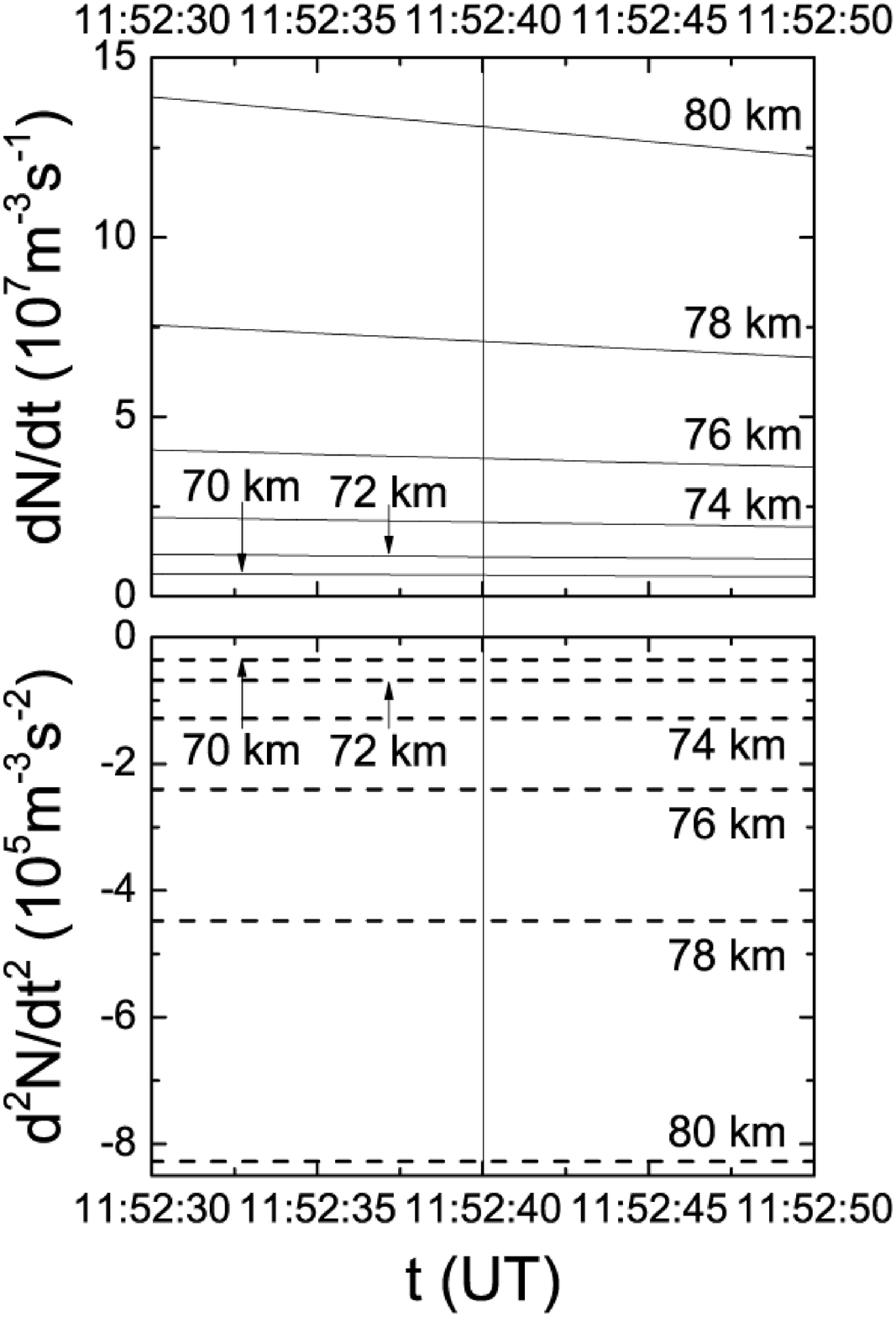}}

\figurecaption{4.}{Altitude dependence of the first and second time derivative of electron density around the time of X radiation maximum ($t_{\rm {I_{max}}}$) at altitudes between 70 km and 80 km}

\section{5. CONCLUSIONS}

In this study we presented a procedure to determine contribution of the solar Ly$\alpha$ line in ionization of the ionospheric D-region at the time of the X radiation maximum intensity during solar X flare events. We applied the obtained theoretical result to the solar X flare of May 5, 2010 while the required input quantities for the electron density are calculated from data obtained experimentally from the D-region monitoring by VLF radio waves. The obtained altitude dependency of the Ly$\alpha$ line contribution in ionization rate shows that the Ly$\alpha$ line contributes less then 0.5\% in the total ionization rate for the considered case. Although the given procedure indicates a larger importance of the Ly$\alpha$ line in ionization processes at lower heights, the corresponding small values of the coefficient $r_{\rm {Ly\alpha}}$ imply the dominant ionization role of the X-radiation spectrum at all considered D-region altitudes.

 \vskip-4.7cm

\centerline{\includegraphics[bb=0 0 629
815,width=\columnwidth, keepaspectratio]{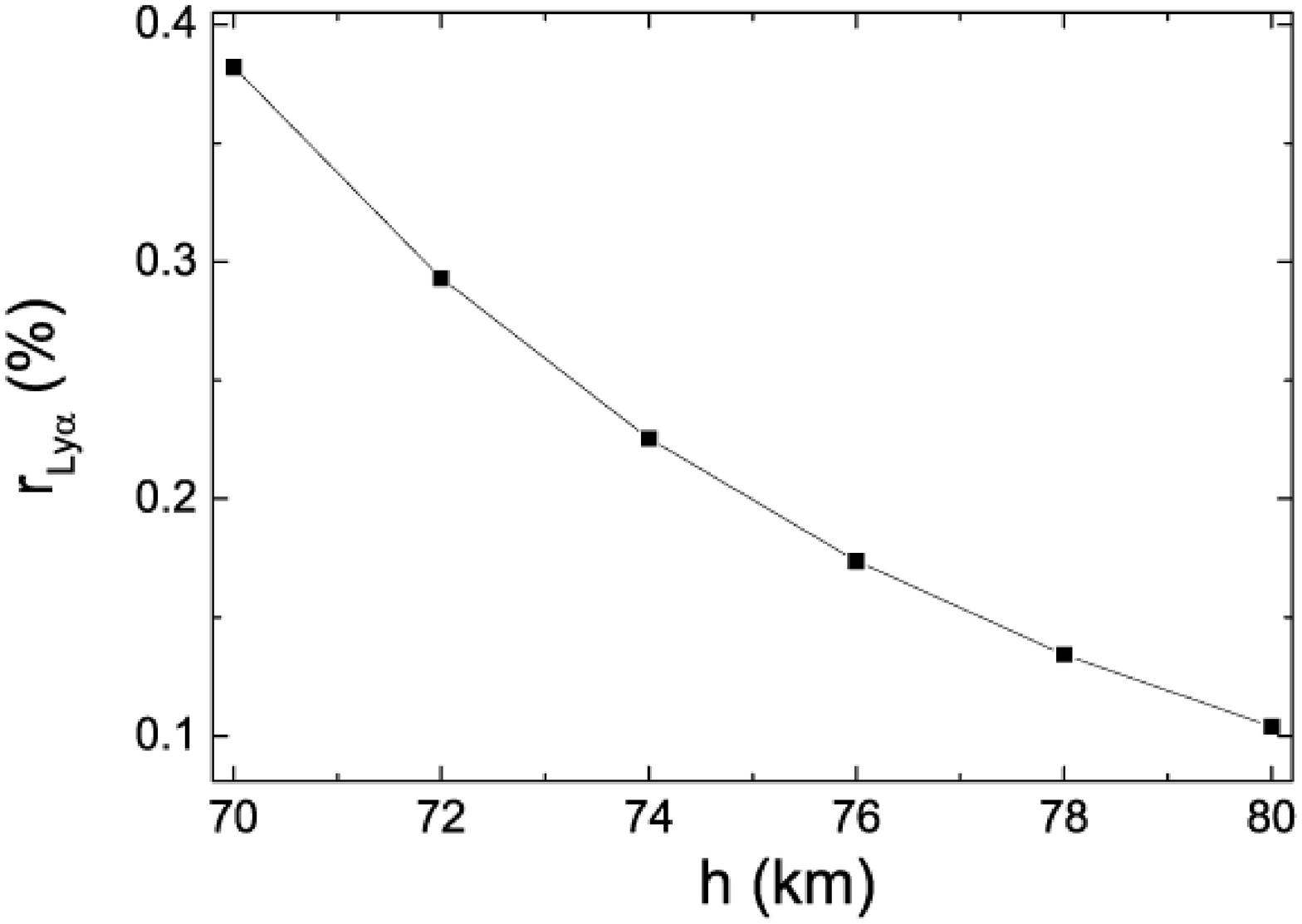}}

\figurecaption{5.}{The relative contribution $r_{\rm {Ly\alpha}}$ of the Ly$\alpha$ radiation in total ionization rate around the time of X radiation maximum ($t_{\rm {I_{max}}}$) at altitudes between 70 km and 80 km}


\acknowledgements{The authors would like to thank the Ministry of Education, Science and Technological Development of the Republic of Serbia for the support of this work within the projects III-44002, 176002 and 176004 and Ministry of Defence within the project VA-TT/OS5/2015.
The data for this paper collected by GOES-14 satellite
is available at NOAA's National Centers for Environmental information (\url{http://satdat.ngdc.noaa.gov/sem/goes/data/new_full/2010/05/goes14/csv/g14_xrs_2s_20100505_20100505.csv}).
The data for this paper related to solar irradiance calculated by Flare Irradiance Spectral Model (FISM) are available on LISIRD site:  \url{http://lasp.colorado.edu/lisird/fism/}. Requests for the VLF
 data used for analysis can be directed to the corresponding author.
}


\references

{Baj\v{c}eti\'{c}}, J., {Nina}, A., \v{C}ade\v{z}, V.~M. and Todorovi\'{c},
  B.~M.: 2015,
\journal{Therm. sci.}, in press (doi: 10.2298/TSCI141223084B).

{Blaunstein}, N. and {Christodoulou}, C.: 2006, {Radio Propagation and Adaptive
  Antennas for Wireless Communication Links: Terrestrial, Atmospheric and
  Ionospheric, John Wiley and Sons, Inc. Hoboken, New Jersey.

{Budden}, K.~G.: 1988,
The Propagation of Radio Waves, Cambridge University Press, Cambridge.

{Chamberlin}, P.~C., {Woods}, T.~N. and {Eparvier}, F.~G.: 2008,
\journal{Space Weather}, \vol{6}, 5001.

Chau, J.~L., R\"{o}ttger, J. and Rapp, M.: 2014,
\journal{J. Atmos. Sol.-Terr. Phy.}, \vol{118}, 113.

{Clilverd}, M.~A., {Thomson}, N.~R. and {Rodger}, C.~J.: 1999,
\journal{Radio Sci.}, \vol{34}, 939.

{Cohen}, M., {Inan}, U. and {Paschal}, E.~W., P.: 2010,
\journal{IEEE T. Geosci. Remote.}, \vol{48}, 3.

{Ferguson}, J.~A.: 1998,
Computer Programs for Assessment of Long-Wavelength Radio
  Communications, Version 2.0, Space and Naval Warfare Systems Center, San Diego.

{Grubor}, D.~P., {{\v S}uli{\'c}}, D.~M. and {{\v Z}igman}, V.: 2005,
\journal{Serb. Astron. J.}, \vol{171}, 29.

{Grubor}, D.~P., {{\v S}uli{\'c}}, D.~M. and {{\v Z}igman}, V.: 2008,
\journal{Ann. Geophys.}, \vol{26}, 1731.

Hayakawa, M., Kasahara, Y., Nakamura, T., Hobara, Y., Rozhnoi, A., Solovieva,
  M. and Molchanov, O.: 2010,
\journal{Journal of Atmospheric and Solar-Terrestrial Physics},
 \vol{72}(13), 982.

Inan, U.~S., Shafer, D.~C., Yip, W.~Y. and Orville, R.~E.: 1988,
\journal{J. Geophys. Res.-Space},
  \vol{93}, 11455.

{Inan}, U.~S., {Cummer}, S.~A. and {Marshall}, R.~A.: 2010,
\journal{J. Geophys. Res.-Space}, \vol{115}, A00E36.

{Jilani}, K., {Mirza}, A.~M. and {Khan}, T.~A.: 2013,
\journal{Astrophysics and Space Science}, \vol{344}, 135.

{Kolarski}, A., {Grubor}, D. and {{\v S}uli{\'c}}, D.: 2011,
\journal{Balt. Astron.}, \vol{20}, 591.

Kolarski, A. and Grubor, D.: 2014,
\journal{Adv. Space Res.}, \vol{53}(11), 1595.

Maurya, A.~K., Phanikumar, D.~V., Singh, R., Kumar, S., Veenadhari, B., Kwak,
  Y.-S., Kumar, A., Singh, A.~K. and Niranjan~Kumar, K.: 2014,
\journal{J. Geophys. Res.-Space}, \vol{119}(10), 8512.

{McEwan}, M. and {Phillips}, F.: 1978,
Chemistry of the Atmosphere, Mir Publishers, Moscow.

{Nina}, A., {{\v C}ade{\v z}}, V., {Sre{\'c}kovi{\'c}}, V.~A. and {{\v
  S}uli{\'c}}, D.: 2011,
\journal{Balt. Astron.}, \vol{20}, 609.

{Nina}, A., {{\v C}ade{\v z}}, V., {Sre{\'c}kovi{\'c}}, V. and {{\v
  S}uli{\'c}}, D. 2012a,
\journal{Nucl. Instrum. Methods in Phys. Res. B},
  \vol{279}, 110.

{Nina}, A., {{\v C}ade{\v z}}, V., {{\v S}uli{\'c}}, D., {Sre{\'c}kovi{\'c}},
  V. and {{\v Z}igman}, V.: 2012b,
\journal{Nucl. Instrum. Methods in Phys. Res. B},
  \vol{279}, 106.

Nina, A. and \v{C}ade\v{z}, V.~M.: 2013,
\journal{Geophys. Res. Lett.}, \vol{40}(18), 4803.

{Nina}, A. and {{\v C}ade{\v z}}: 2014,
\journal{Adv. Space Res.}, \vol{54}(7), 1276.

{Price}, C., {Yair}, Y. and {Asfur}, M.: 2007,
\journal{Geophys. Res. Lett.}, \vol{34}, 9805.

{Ratcliffe}, J.~A.: 1972,
{An Introduction to the Ionosphere and the Magnetosphere}, Cambridge University Press, Cambridge.

Raulin, J.-P., Trottet, G., Kretzschmar, M., Macotela, E. L., Pacini, A., Bertoni, F. C. P. and
Dammasch, I. E.: 2013, \journal{J. Geophys. Res.-Space}, \vol{118}, 570.

Schmitter, E.~D.: 2013,
\journal{Ann. Geophys.}, \vol{31}(4), 765.

{Singh}, R., {Veenadhari}, B., {Maurya}, A.~K., {Cohen}, M.~B., {Kumar}, S.,
  {Selvakumaran}, R., {Pant}, P., {Singh}, A.~K. and {Inan}, U.~S.: 2011,
\journal{J. Geophys. Res.-Space}, \vol{116}, 10301.

Singh, A.~K., Singh, A., Singh, R. and Singh, R.: 2014,
\journal{Astrophys. Space Sci.}, \vol{350}(1), 1.

{Strelnikova}, I. and {Rapp}, M.: 2010,
\journal{Adv. Space Res.}, \vol{45}(2), 247.

{\v{S}uli\'{c}}, D. and {Sre\'{c}kovi\'{c}}, V.~A.: 2014,
\journal{Serb. Astron. J.}, \vol{188}, 45.

Ushio, T., Wu, T. and Yoshida, S.: 2014,
\journal{Atmos. Res.}, \vol{154}, 89.

{{\v Z}igman}, V., {Grubor}, D. and {{\v S}uli{\'c}}, D.: 2007,
\journal{J. Atmos. Sol.-Terr. Phy.},
  \vol{69}, 775.

{{\v Z}igman}, V., {Kudela}, K. and {Grubor}, D.: 2014,
\journal{Adv. Space Res.},
  \vol{53}(5), 763.

{Wait}, J.~R. and {Spies}, K.~P.: 1964,
Characteristics of the Earth-ionosphere waveguide for VLF radio
  waves, NBS Technical Note 300, National
Bureau of Standards, Boulder, CO.

Zhang, X. and Tang, L.: 2015,
\journal{Ann. Geophys.}, \vol{33}(1), 137.}

%
%
%
%
%
%
%
%
%
%
%
%

\endreferences

\end{multicols}

\vfill\eject

{\ }



\naslov{Udeo emisije u liniji {{$\rrm {\mathrm{L\lowercase{y}}}\alpha$}} Sun{\CH}evog vodonika u ukupnom stepenu jonizacije jonosferske D-oblasti tokom maksimuma Sun{\CH}eve erupcije u H-podru{\CH}ju}


\authors{
        A. Nina$^{1}$,
        V. M. \v{C}ade\v{z}$^{2}$
        and J. Baj\v{c}eti\'{c}$^{3}$}

\vskip3mm


\address{$^1$Insttute of Physics,
University of Belgrade\break Pregrevica 118, 11080 Belgrade,
Serbia}

\Email{sandrast}{ipb.ac.rs}

\address{$^2$Astronomical Observatory, Volgina 7, 11060 Belgrade 38, Serbia}

\Email{vcadez}{aob.rs}

\address{$^2$Department of Telecommunications and Information Science, University of Defence, Military Academy, Generala Pavla Juri\v{s}i\'{c}a \v{S}turma 33, 11000 Belgrade, Serbia}

\Email{bajce05}{gmail.com}

%
%
%
%
%
%

\vskip.7cm


\centerline{UDK \udc}


\centerline{\rit Originalni nuqni rad}

\vskip.7cm

\begin{multicols}{2}
{


{\rrm Zra{\ch}e{\nj}e u }Ly$\alpha$ {\rrm liniji se mo{\zz}e smatrati dominantnim izvorom jonizacionih procesa u neporeme{\cc}enoj jonosferskoj D-oblasti na visinama iznad oko 70 km. Me{\dj}utim, intenzivni upadi zra{\ch}e{\nj}a u nekim drugim energijskim podru{\ch}jima mogu tako{\dj}e zna{\ch}ajno uticati na brzinu jonizacije. U ovom radu predstav{\lj}amo studiju o udelu} Ly$\alpha$ {\rrm zra{\ch}e{\nj}a u brzini jonizacije tokom perioda kada je jonosfera poreme{\cc}ena Sun{\ch}evim erupcijama u H-oblasti. Dati su odgovaraju{\cc}i analiti{\ch}ki izrazi i ura{\dj}eni prora{\ch}uni i numeri{\ch}ke simulacije za nisku jonosferu kori{\sh}{\cc}e{\nj}em podataka prikup{\lj}enih} VLF {\rrm prijemnikom koji se nalazi u Srbiji za} VLF {\rrm radio-signal emitovan} DHO {\rrm predajnikom u Nema{\ch}koj.
}
}


%
\end{multicols}

\end{document}